\documentclass[a4paper,aps,prl,superscriptaddress,column,reprint]{revtex4-1} 
\usepackage{graphicx}
\usepackage[small,justification=justified]{caption} 
\usepackage{braket}
\usepackage[utf8]{inputenc}
\usepackage{hyperref}
\usepackage{amsmath} 
\usepackage{amssymb}  
\usepackage{amstext}  
\usepackage{graphicx}
\usepackage{amssymb}
\usepackage{color}
\usepackage{subcaption}

\begin{document}

\title{Kinks and Nanofriction: Structural Phases in Few-Atom Chains}

\author{Dorian A. Gangloff}
\email[Electronic address: ]{dag50@cam.ac.uk}
\affiliation{Cavendish Laboratory, University of Cambridge, JJ Thomson Avenue, Cambridge CB3 0HE, United Kingdom}

\author{Alexei Bylinskii}
\affiliation{Department of Chemistry and Chemical Biology and Department of Physics, Harvard University, Cambridge MA, 02138, USA}

\author{Vladan Vuleti\'c}
\email[Electronic address: ]{vuletic@mit.edu}
\affiliation{Department of Physics, and Research Laboratory of Electronics, Massachusetts Institute of Technology, Cambridge MA, 02139, USA}

\begin{abstract}
The frictional dynamics of interacting surfaces under forced translation are critically dependent on lattice commensurability. Performing experiments in a trapped-ion friction emulator, we observe two  distinct structural and frictional phases: a commensurate high-friction phase where the ions stick-slip simultaneously over the lattice, and an incommensurate low-friction phase where the propagation of a kink breaks that simultaneity. We experimentally track the kink's propagation with atom-by-atom and sub-lattice site resolution, and show that its velocity increases with commensurability. Our results elucidate the commensurate-incommensurate transition and the connection between the appearance of kinks and the reduction of friction in a finite system, with important consequences for controlling friction at nanocontacts.
\end{abstract}

\maketitle

Commensurability at the interface between two atomically smooth, elastic surfaces can fundamentally alter the energetic cost of their forced relative motion \cite{Vanossi2013,Krylov2014}. Commensurate surfaces experience the largest sticking forces, and thus the most discontinuous form of motion: stick-slip friction \cite{Urbakh2004}. At sufficient mismatch between the two surface lattices, the interface develops defects distributed over a collection of atoms -- kink solitons -- that result in an incommensurate phase with smoother surface translation, reduced energy barriers, and reduced friction \cite{Braun1990b}. Thus the appearance of kinks, that requires finite lattice mismatch, marks the commensurate-incommensurate (C-I) transition \cite{Bak1982,Pokrovsky1979,Coppersmith1982,Buchler2003}.

In the solid state, two-dimensional versions of the C-I transition were observed at the interface between krypton monolayers and graphite \cite{Fain1980}, and between graphene and hexagonal boron nitride \cite{Woods2014}. There, rotation between two surfaces introduced a lattice mismatch and the transition between a commensurate and an incommensurate phase. However, a direct experimental link between the C-I transition and the appearance of a finite kink density, supported by an atomistic calculation of the energy landscape \cite{Dezerald2014}, is challenging in such systems. 

Typically, the Frenkel-Kontorova model \cite{BraunKivsharFKbook} is used to describe the physics of commensurability of extended interfaces. In this model, the motion of kinks occurs within a potential landscape, the Peierls-Nabarro (PN) potential. The concepts of the PN potential and kinks naturally extend to small systems, as for example in the diffusion of epitaxial islands and surface adsorbates \cite{Markov1980,Braun1990b,Willis1986,Kwasniewski1999}. Small systems offer an opportunity to study the formation of kinks under conditions where a direct link between concepts and experimental observations can be established.

Friction emulators with synthetic interfaces consisting of colloidal particles \cite{Bohlein2012} or cold trapped ions \cite{Bylinskii2015} have improved our understanding of fundamental surface science, owing to in-situ tuning of fundamental parameters, and to the ability to image individual particles. Such emulators \cite{Benassi2011,Mandelli2013,Mandelli2015c} have been used to observe the Aubry transition \cite{Bylinskii2016a,Kiethe2017a,Brazda2018} and kink transport \cite{Bohlein2012,Brox2017}. Cold trapped-ion systems \cite{Bloch2005,Enderlein2012,Linnet2012,Karpa2013a} have been used to study fundamental aspects of atomistic friction, such as mismatch-induced lubricity \cite{Bylinskii2015}, the temperature and velocity dependence of friction \cite{Gangloff2015b}, and multislip friction \cite{Counts2017a}.

\begin{figure*}[hbpt]
  	\begin{center}
   \includegraphics*[width=\textwidth]{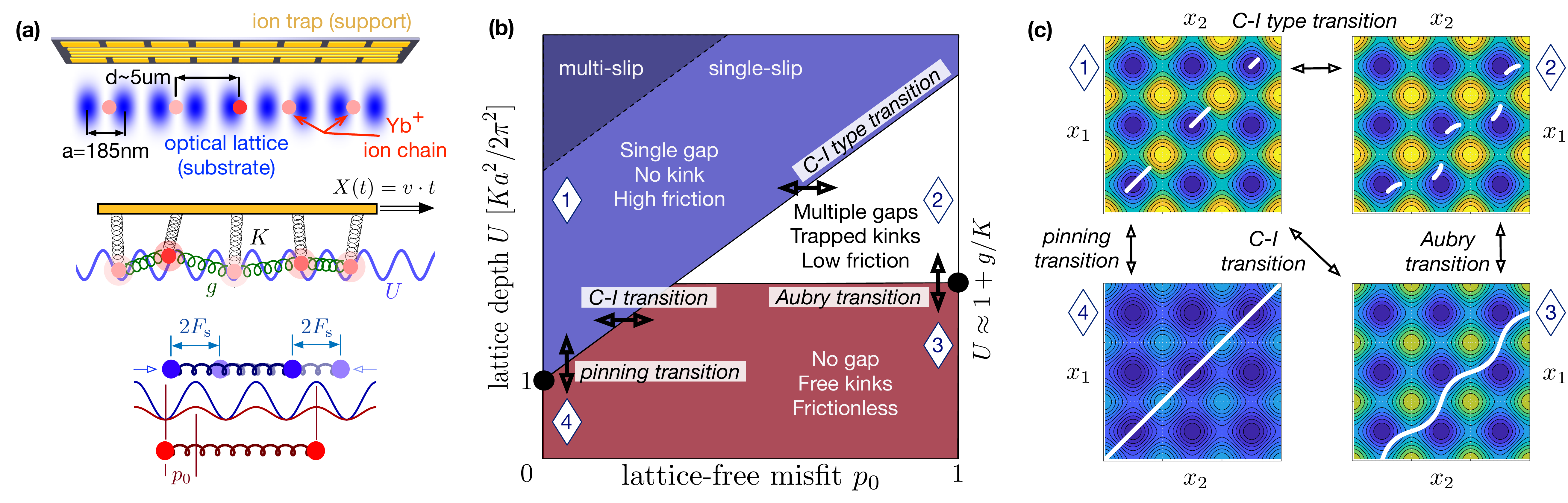}
             \centering
   \caption{\textbf{Ion-crystal friction emulator for the study of structural phases and kinks.} 
   \textbf{(a)} Our friction emulator \cite{Bylinskii2015}: a one-dimensional crystal of ions and an optical lattice \cite{Karpa2013a,Cetina2013}. The crystal is displaced across the lattice with ion-trap electric fields. Each ion's interaction with the support and with its neighbor are captured by spring constants $K$ and $g$, respectively \cite{Weiss1996,Bylinskii2016a,SuppInfo}. The lattice-free misfit $p_0$ is measured at low lattice depth $U$ (red). At high lattice depth (blue), displacement hysteresis is equivalent to friction $F_\text{s}$. \textbf{(b)} Schematic structural phase diagram for a two-ion chain in the ($p_0,U$) plane. Arrows highlight transitions of interest. 
   \textbf{(c)} Adiabatic path (white curve) of a two-ion chain through the lattice energy landscape (minima in blue, maxima in yellow) as the trap is translated, vs ion positions $x_1,x_2$, shown for points of interest in the ($p_0,U$) plane.}
       \label{fig:Fig1}
       \end{center}
\end{figure*}

In this Letter, we use a trapped-ion friction emulator to experimentally observe a structural phase transition captured by the appearance of kinks: the few-atom analogue of the C-I transition. We observe the stick-slip dynamics of the chain atom by atom, find that a critical degree of commensurability is required for kinks to form, and tie their appearance to a reduction in the observed friction.

Our emulator \cite{Bylinskii2015,Karpa2013a,Cetina2013} consists of a self-organized one-dimensional Coulomb crystal of $N$ laser-cooled $^{174}$Yb$^+$ ions in a linear Paul trap \cite{Leibfried2003a}, sliding over a periodic optical potential generated by a standing wave of light (Fig.~\ref{fig:Fig1}a). A translation of the Paul trap's potential minimum $X_t$ with respect to the optical lattice transports the ion crystal at adjustable speed $v_t$ over the periodic potential. The position of each ion relative to the optical lattice is tracked with subwavelength resolution via their fluorescence \cite{SuppInfo}, allowing us to measure each ion's hysteresis loops \cite{Bylinskii2015} and to reconstruct a kink traveling through the ion chain. Continuous laser cooling of the ion chain removes heat generated by friction, while the effects of finite temperature are reduced by performing experiments at sufficiently high translation speed \cite{Gangloff2015b}.

Although the intrinsic ion spacing $d$ of a few micrometers is not uniform along the chain, it can be controlled with nanometer precision by adjusting the Paul trap harmonic potential. A deviation of $d ($mod $a)$ from $a$ introduces an average lattice-free misfit,
\begin{multline*}
    p_0/2 =     \left\{ \begin{array}{rcl}
          d/a \mbox{ mod } 1 & \mbox{for}
         & 0 \leq d \mbox{ mod } a < 0.5a \\
         1 - d/a \mbox{ mod } 1 & \mbox{for} & 
         0.5a \leq d \mbox{ mod } a < a
                \end{array}\right. \mbox{,}
\end{multline*}
between the ion crystal and the lattice with period $a=185$~nm (Fig.~\ref{fig:Fig1}a). The chain is in-registry with the substrate when the average ion spacing $d$ is an integer multiple of the lattice spacing $a$, $d/a ($mod $1) = 0$, corresponding to no misfit, $p_0 = 0$. A departure from this configuration introduces a misfit $p_0>0$ whose value can be tuned continuously up to $p_0=1$ for $N=2$, and up to $p_0 \approx 0.8$ for $N=5$. A non-zero misfit can cause up to $Np_0/2$  kinks -- i.e. the kink density $p_\text{k}$ is at most $p_0$ \cite{Bak1982}. Here we tune the lattice-free misfit $p_0$ from 0 to 1 for $N=2$, and from 0 to 0.4 for $N=5$, which introduces at most one kink.

Several structural phases are expected as a function of the lattice-free misfit $p_0$ and the lattice depth $U$, as shown schematically in Fig.~\ref{fig:Fig1}b. For large misfit $p_0 \sim 1$, emulating the maximally incommensurate case for infinite chains, the Aubry transition \cite{Aubry1983a} at critical lattice depth $U_c$ takes the system from a frictionless phase (for $U<U_c$) to a pinned phase (for $U>U_c$), as observed previously in our finite system \cite{Bylinskii2016a}. By contrast, the C-I transition occurs as a function of the misfit $p_0$, and can be identified by the appearance of kinks \cite{Bak1982}. 

Figure~\ref{fig:Fig1}c illustrates these important transitions for a finite system in the limiting case $N=2$. For a sufficiently commensurate arrangement $p_0 \sim 0$ (Fig.~\ref{fig:Fig1}b, diamond 1), as the trap position $X_t$ is translated over the lattice potential, the chain is pinned into a configuration where the atoms are in-registry with the lattice, and where the forced translation of the chain creates sudden transitions towards the same stable configuration shifted by one lattice site (Fig.~\ref{fig:Fig1}c, diamond 1). The stick-slip events of all atoms in the chain are synchronous and result in the largest friction force -- a regime well described by Prandtl-Tomlinson physics \cite{Vanossi2013}. As the misfit $p_0$ is increased, the different atoms experience different lattice forces -- a regime better described by Frenkel-Kontorova physics \cite{BraunKivsharFKbook}. At a critical value $p_0 \geq p_\text{c}(U)$, it becomes energetically favorable under forced translation for some of the atoms in the chain to slip to the next lattice site, while the rest of the chain remains in its initial lattice site (Fig.~\ref{fig:Fig1}c, diamond 2). This event defines a configurational change in the chain to a new metastable state that contains a kink defect \cite{Frank1949,Braun1990b} (Fig.~\ref{fig:Fig1}b, diamond 2). The appearance of a kink embodies a C-I transition \cite{Bak1982} for a finite system, although the metastable kink state is not truly incommensurate \cite{Sharma1984,Joos1983,Braun1990b}. In this configuration, the PN barriers are finite: kinks are trapped, the translation of the chain is still dominated by stick-slip events, and the adiabatic trajectory has gaps and is thus not analytic. Below a critical value of the lattice depth ($U<U_c$), marking the Aubry transition, the PN barriers disappear: the movement of kinks is free, the translation of the chain over the substrate is continuous, the function defining its adiabatic trajectory is analytic, and friction disappears (Fig.~\ref{fig:Fig1}b,c, diamond 3). PN barriers also disappear at $p_0\sim 0$ when the trap force exceeds the maximal substrate force \cite{Socoliuc2004}, $U<Ka^2/2\pi^2$ (Fig.~\ref{fig:Fig1}b, diamond 4).

The kink density $p_\text{k}$ is simply the misfit measured in the presence of the lattice \cite{Bak1982}, as it quantifies the deviation per particle from a commensurate arrangement. At a critical value of $p_0$ that depends on the lattice depth $U$, the kink density $p_\text{k}$ undergoes a transition from $p_\text{k}=0$ to $p_\text{k}>0$, as shown in Fig.~2a for an infinite Frenkel-Kontorova chain. This critical point $p_\text{c}(U)$, beyond which $p_\text{k}$ quickly converges to $p_0$, indicates the C-I transition \cite{Bak1982}.

\begin{figure}[t]
  	\begin{center}
   \includegraphics*[width=\columnwidth]{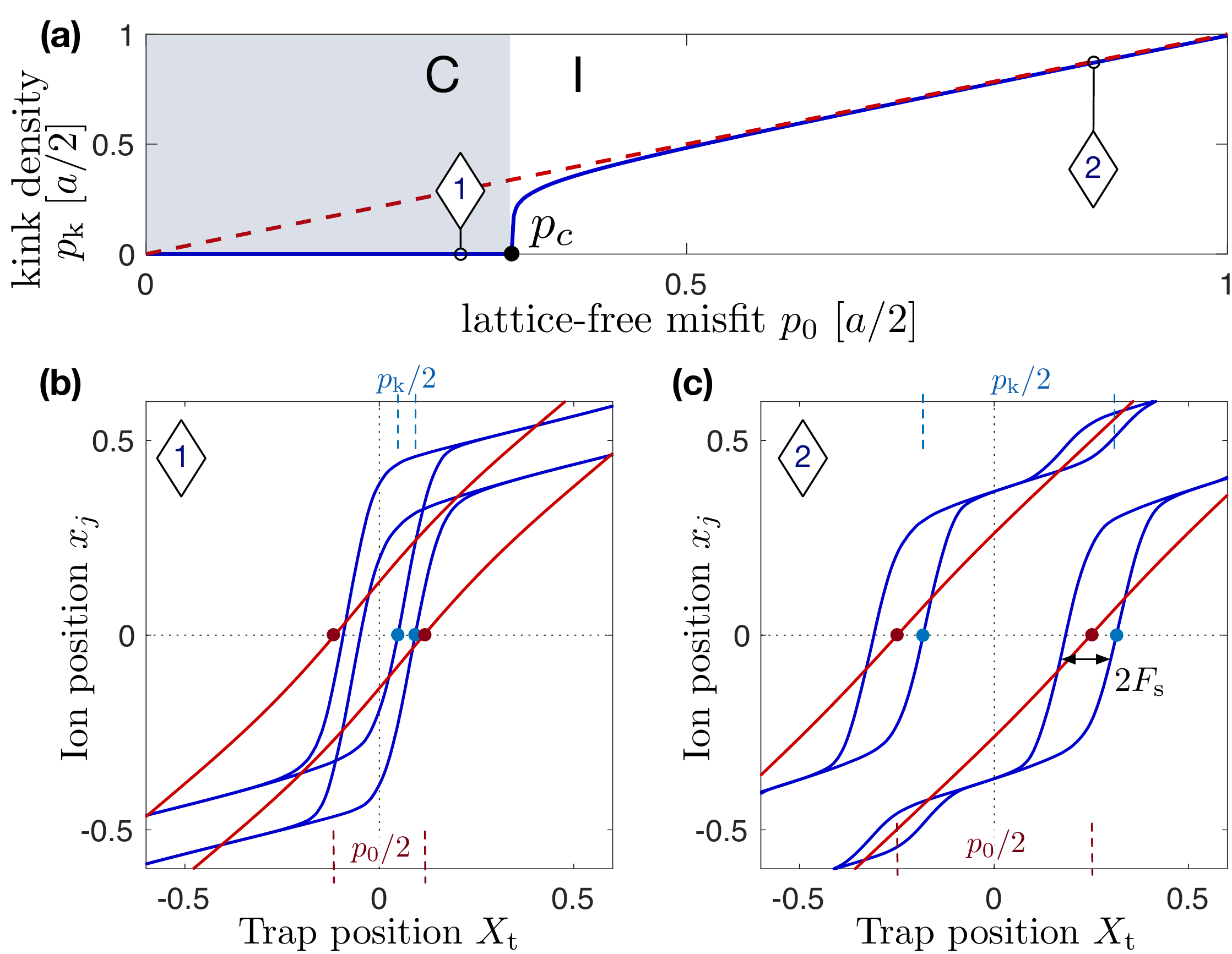}
          \centering
          \caption{\textbf{The C-I transition via kink density and hysteresis} 
             \textbf{(a)} Kink density $p_\text{k}$ vs lattice-free misfit $p_0$ in the infinite one-dimensional Frenkel-Kontorova model, showing the C-I transition \cite{Bak1982} at the critical misfit $p_\text{c}$ separating the commensurate (C) and incommensurate (I) phases.
             \textbf{(b-c)} Simulated ion positions $x_j$ vs trap position $X_t$ in their lattice-free position (red) and lattice-perturbed position (blue), where forward and backward translation results in static friction $F_s$. The difference in trap positions at which the ions pass lattice maxima $x_j=0$ corresponds to the misfit $p_0$ (red) and the kink density $p_\text{k}$ (blue). (b) $p_0 = 0.3$. (c) $p_0 = 0.8$.}
       \label{fig:Fig2}
       \end{center}
\end{figure}

To observe this transition in our finite chain, we evaluate the trap positions $X_t^{(j)}$ at which ion $j$, under forced translation, passes a lattice maximum ($x_j = 0$) in relation to its neighbor $j+1$. The difference $X_t^{(j+1)} - X_t^{(j)}$ captures whether slip events are asynchronous, and therefore whether a kink is present. Figure~\ref{fig:Fig2}b,c shows the simulated trajectories $x_1,x_2$ of two ions against trap position $X_t$ \cite{SuppInfo}. The red curves are the lattice-free ($U=0$) trajectories: the horizontal distance between them is the lattice-free misfit $p_0$, and overlapping curves ($p_0=0$) show the chain is commensurate with the lattice. The blue curves are trajectories in the presence of a deep lattice ($U>U_\text{c}$): they exhibit hysteresis -- and therefore friction -- corresponding to a gapped adiabatic trajectory (Fig.~\ref{fig:Fig1}c). At misfits below the C-I transition (Fig.~\ref{fig:Fig2}b), the trajectory has a single gap (per lattice period), and the hysteresis loops of the two ions overlap almost completely. This corresponds to synchronous slip events characterized by a kink density $p_\text{k} \sim 0$. At large misfit (Fig.~\ref{fig:Fig2}c), the hysteresis loops exhibit no overlap, corresponding to asynchronous slip events and to the maximum possible kink density $p_\text{k} \sim p_0$. The doubling of gaps (per lattice period) in the adiabatic trajectory, and the consequent doubling of hysteresis loops, can be understood as two transitions: the entry of a kink at one end of the chain, and its exit at the other end. The intermediate state is a distinct, metastable configuration of the chain containing a trapped kink \cite{Braun1990b}.

\begin{figure}[hbpt]
   \includegraphics*[width=\columnwidth]{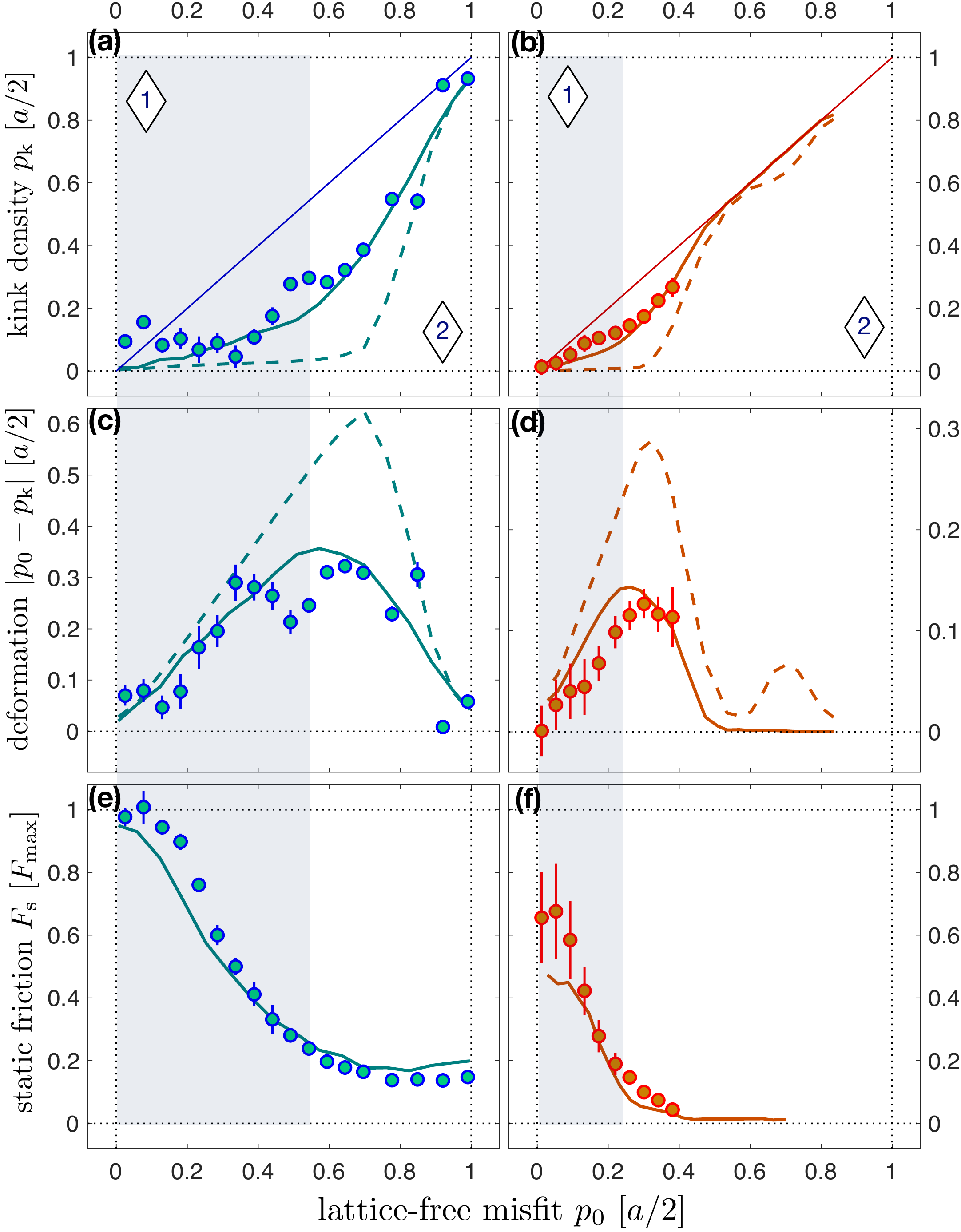}
\caption{\textbf{Kinks in a two-ion chain (left column) and a five-ion chain (right column).} Kink density $p_\text{k}$ (a,b), deformation $|p_0-p_\text{k}|$ (c,d), and static friction force $F_s$ (e,f), measured vs lattice-free misfit $p_0$ for $N=2$ (blue, a, c, e) and $N=5$ (red, b,d,f). Lines in (a,b) are the kink density $p_\text{k} = p_0$ for $U=0$. Solid curves in (a-d) are numerical simulations for parameters matching those of the data shown, $kT/U=0.2$ (a) and $kT/U=0.5$ (b). Dashed curves are $T=0$ simulations. The critical misfit $p_\text{c}(U)$ is the misfit $p_0$ that maximises $p_0-p_\text{k}$: $p_\text{c}(U)=0.55$ ($N=2$) and $p_\text{c}(U)=0.25$ ($N=5$). Numbered diamonds match Fig.~1 points of interest. $F_\text{max}$ is the maximal friction in the single-slip regime\cite{Gangloff2015b,Counts2017a}.}
\label{fig:Fig3}
\end{figure}

Figure 3 shows the measured kink density $p_\text{k}$ and static friction force $F_\text{s}$ as a function of the lattice-free misfit $p_0$ for a two- and a five-ion chain. The  trap positions $X_t^{(j)}$ at which ion $j$ passes a lattice antinode (the "slip" positions) appear as fluorescence peaks in our experiment \cite{Bylinskii2015}. A linear fit of the slip positions $X_t^{(j)}$ vs $j$ gives $p_\text{k}$, while the difference between forward and backward slips gives $F_\text{s}$. The lines $p_\text{k} = p_0$ in Fig.~\ref{fig:Fig3}a,b represent the expected behaviour for $U \ll U_c$, when ion movement is unperturbed by the lattice. A finite lattice depth ($U>U_c$) pins the ions to lattice minima even for $p_0>0$, causing the ions to slip more synchronously, and thereby lowering the kink density to $p_\text{k} \ll p_0$. Indeed, measurements of $p_\text{k}$ as a function of $p_0$ agree qualitatively with the critical behaviour expected from the infinite-chain Frenkel-Kontorova physics (Fig.~\ref{fig:Fig2}a): for $p_0$ less than a critical value $p_\text{c}$, $p_\text{k}$ stays close to $0$ (synchronous slipping), while above that critical value, $p_\text{k}$ ascends rapidly back towards $p_0$ (asynchronous slipping). The transition is highlighted by the deformation $|p_0-p_\text{k}|$, which reaches a maximum at the critical point $p_\text{c}(U)$, as shown in Fig.~\ref{fig:Fig3}c,d.

The transition is smoother than expected from an infinite Frenkel-Kontorova chain, which we attribute to finite-temperature and finite-size effects. Simulations of a finite chain at temperature $T$ \cite{SuppInfo}, shown as solid curves in Fig.~\ref{fig:Fig3}a-d, are in good agreement with our data using $kT/U = 0.2$ for $N=2$ and $kT/U=0.5$ for $N=5$. By comparison, zero-temperature simulations of the same systems, shown as dashed curves, exhibit a sharper transition that occurs at a higher value of the misfit $p_0$. Temperature effectively reduces the PN barriers that cause stick-slip motion, making it more likely for asynchronous slips to occur, and thus increasing $p_\text{k}$ relative to its zero-temperature value. The additional step structure for $N=5$ reflects the fact that, for $N>3$, intermediate values of the misfit correspond to the system entering higher-order commensurate phases. As $N$ grows the number of steps grows until, for $N=\infty$, this structure constitutes a Devil's staircase \cite{Bak1982,BraunKivsharFKbook}.

The transition points $p_\text{c}\sim0.55$ for $N=2$ and $p_\text{c} \sim 0.25$ for $N=5$, associated with the appearance of kinks, also delineate two frictional phases, as shown in Fig.~\ref{fig:Fig3}e,f. In the region $p_0 \lesssim p_\text{c}$ (Fig.~\ref{fig:Fig1}, diamond 1) where $p_\text{k} \sim 0$ and the slips are synchronous, friction is high, while in the region $p_0 \gtrsim p_\text{c}$ (Fig.~1, diamond 2) where $p_\text{k} \sim p_0$ and the slips are asynchronous, friction is low. The transition point $p_\text{c}$ also marks a reduction in the dependence of friction $F_s$ on the misfit $p_0$, thus confirming that the presence of a kink defect changes the frictional response of the system. 

\begin{figure}[hbpt]
\centering
      \includegraphics[width=\columnwidth]{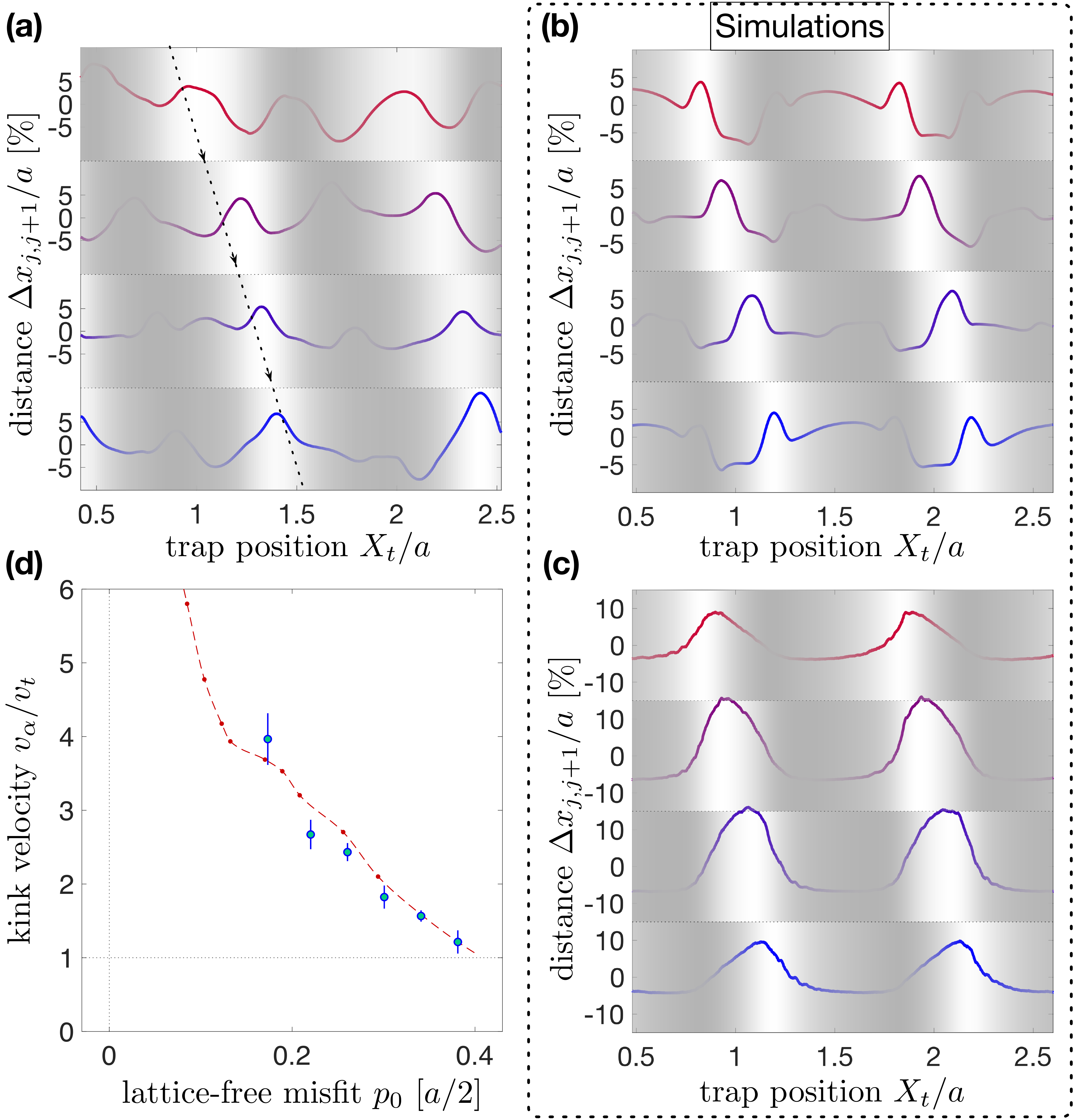}
   \caption{\textbf{Propagation of a kink in a five-ion chain} \textbf{(a)} The distance $(\Delta x_{j,j+1} = (x_j - x_{j+1})) (\text{mod } a)$ between ion $j$ and its nearest neighbor $j+1$ (with mean subtracted) vs trap position $X_t$. From top to botttom, $j=1,...,4$. Dashed line tracks propagation of the kink maximum across the ion pairs. Grey shading indicates regions where fluorescence is low, and position is extracted with lower confidence. \textbf{(b)} Numerical simulations for parameters matching those of data in (a). $\Delta x_{j,j+1}$ is reconstructed from a simulated fluorescence signal. \textbf{(c)} $\Delta x_{j,j+1}$ from simulated position curves\cite{SuppInfo}. \textbf{(d)} Kink velocity $v_\alpha$, normalized by trap velocity $v_t$, against lattice-free misfit $p_0$. The dashed line is from simulation.}
       \label{fig:Fig4}
\end{figure}

The incommensurate phase ($p_0 > p_\text{c}$) defined by asynchronous slipping must exhibit a traveling kink, whose direct measurement is possible in a sufficiently long chain. We confirm this from the position of all ions in a five-ion chain as they traverse the optical lattice. Using their known position-dependent fluorescence across the lattice, which follows a $\sin^4(x_j)$ function, we reconstruct each ion's position from its fluorescence \cite{SuppInfo,Gangloff2016a}. We obtain pair-wise the neighbor distance $\Delta x_{j,j+1}$ ($j=1,...,4$), and track a compression and extension in relative position within each ion pair as the background trap is translated (Fig.~\ref{fig:Fig4}a). Grey shading denotes low fluorescence, and therefore limited reconstruction fidelity \cite{SuppInfo}.

Clear oscillations in the neighbor distances (Fig.~4a) reveal a density wave travelling through the discrete system: a kink enters the chain at its free end, travels through it and exits at the other free end, causing a staggered signal across pairs which repeats at the lattice period $a$ as the Paul trap translates the chain. For a given pair, the distance $\Delta x_{j,j+1}$ reaches a maximum when the kink profile is centered on that pair.  This analysis is supported by finite-temperature numerical simulations of our experiment with a five-ion chain (Fig.~4b,c). In Fig.~4b, the signal $\Delta x_{j,j+1}$ is reconstructed from simulated fluorescence curves,
processed identically with our data. In Fig.~4c, $\Delta x_{j,j+1}$ is calculated from the simulated position curves -- which, unlike in the experiments, is accessible in simulation. While the curves reconstructed from fluorescence (Fig.~4b) are clearly deformed, and exhibit the same erroneous deformations as our data in the low-fluorescence regions (grey area), they are a good approximation to the true distance (Fig.~4c).

Tracking the positions of the kink maximum across the pairs as the trap is translated (Fig.~\ref{fig:Fig4}a) yields the velocity $v_\alpha$ of the traveling kink. This is summarized in Fig.~\ref{fig:Fig4}d, where the kink velocity is normalized by the trap translation velocity $v_t = dX_t/dt$. The dashed curve is obtained from our numerical simulations of the fluorescence signal. Our data reproduce the salient feature in the simulation: the kink velocity increases with decreasing misfit $p_0$ as expected from a system approaching a commensurate phase with increasingly synchronous slips. Below $p_0 \approx 0.18$, the staggered kink signal fades in both data and simulation, in agreement with a transition to the commensurate phase.

In summary, we observe atom-by-atom the appearance of a kink in a finite system and connect it to a structural phase transition between a commensurate phase, where friction is high, and an incommensurate phase, where friction is reduced. This work could enable the study of interacting topological defects and frustration at a nanocontact\cite{BraunKivsharFKbook}. Furthermore, quantum tunnelling of ions through lattice barriers, in principle realizable in our system\cite{Bylinskii2016,Gangloff2016a,Zanca2018}, could introduce a quantum-mechanical picture of kinks with relevance at the nanoscale and at cold surfaces.

\begin{acknowledgments}
This work was supported in part by the NSF, the NSF CUA, and the MURI program through ONR. D.G. acknowledges support from a St John's College Research Fellowship.
\end{acknowledgments}

\bibliographystyle{unsrt}

\end{document}